\newcommand{\rem}[1]{}
\documentclass[prd,twocolumn,showpacs,preprintnumbers,amsmath,amssymb]{revtex4-2}
\usepackage{amsfonts,amssymb,amsmath,mathrsfs}
\usepackage{url}
\usepackage{graphicx}
\newtheorem{thrm}{Theorem}%[section]

\newtheorem{prop}[thrm]{Proposition}

\begin{document}
%%%%%%%%%%%%%%%%%%%%%%%%%%%%%%%%%%%%%%%%%%%%%%%%%%%%%%%%%

\title[TOV equation in conformal Killing gravity]{Tolman-Oppenheimer-Volkoff
equation\\ and static spheres 
 in Conformal Killing gravity }

\author{Carlo Alberto Mantica}
\author{Luca Guido Molinari} 
\affiliation{Physics Department Aldo Pontremoli,
Universit\`a degli Studi di Milano and I.N.F.N. sezione di Milano,
Via Celoria 16, 20133 Milano, Italy.}
%OrcId: 0000-0001-5638-8655, 0000-0002-5023-787X}
\email{carlo.mantica@mi.infn.it, luca.molinari@unimi.it}

\begin{abstract} 
We derive the analog of the Tolman-Oppenheimer-Volkoff equation in conformal Killing gravity in a static spherically symmetric spacetime, sourced by anisotropic fluid matter. It differs from the original equation by new `dark terms' associated to a conformal Killing tensor. The formulation of gravity as an Einstein equation augmented by a conserved conformal Killing tensor enables to implement the junction conditions for a sphere of anisotropic fluid 
with the conformal Killing vacuum. The equations are solved for a perfect fluid sphere with uniform matter density, in the Harada vacuum. The extension of Buchdahl's equation for the critical radius - mass density is obtained. 
\end{abstract}
\date{25 Jan 2025}

%\subjclass[2010]{83C20 %Classes of solutions; algebraically special solutions, metrics with symmetries for problems in general relativity and gravitational theory 
%(Primary), 
%83C55, %Macroscopic interaction of the gravitational field with matter (hydrodynamics, etc.)
%83D05 %Relativistic gravitational theories other than Einstein’s, including asymmetric field theories
%(Secondary)}
%\keywords{Conformal Killing gravity; conformal Killing tensor; Robertson-Walker space-time}

\maketitle

%\keywords{\textit{Tolman-Oppenheimer-Volkoff equation, conformal Killing gravity, extended theories of gravity, conformal
%Killing tensor, static spherically symmetric spacetimes, junction conditions, Buchdahl limit}}

\section{Introduction}
The Tolman-Oppenheimer-Volkoff (TOV) equation appeared in the paper on neutron star cores by Oppenheimer and Volkoff \cite{OppVolk39}. It descends from Tolman's equations for static perfect fluid spheres in General Relativity, 
published with solutions in 1939 in the same issue of the Physical Review \cite{Tolman39}, and stated in 1934 in his classic book \cite{Tolman34}. 
The equation relates the pressure $p (r)$, the mass density $\rho (r)$ and the inner mass $M(r) = 4\pi \int_0^r r'^2 \rho (r') dr'$ of a perfect fluid  in static spherical symmetry:
\begin{align}
\frac{dp}{dr}\left [\frac{1}{\kappa_G}-\frac{M(r)c^2}{4\pi r} \right ]= - \frac{p+\rho c^2}{2}\left [pr +\frac{M(r)c^2}{4\pi r^2}  \right ]. \label{TOV_GR}
\end{align}
$\kappa_G = 8\pi G/c^4$ is the gravitational constant.

If matter fills a sphere with surface area $4\pi R^2$, the exterior vacuum is Schwarzschild, with the mass of the star $M(R)$ as unique parameter. This is a consequence of Birkhoff's theorem (1923): spherically symmetric vacuum solutions of the Einstein equations are static. \\
The TOV equation has been studied in several settings as stars, compact objects and black holes, in General Relativity \cite{Zurek84}\cite{Gorini08}\cite{Gorini09}\cite{Carloni18a}\cite{Carloni18b}\cite{Anastopoulos21} and in modified 
gravity \cite{Momeni15}. The extension to charged matter has been considered in \cite{Bekenstein71}\cite{Ray03}\cite{Arbanil13}.

In this note we derive the analog of the TOV equation 
and study the uniform static sphere in conformal Killing gravity (CKG), an extension of Einstein's gravity
proposed by J. Harada in 2023 \cite{Harada23,Harada23b}:%
\begin{equation}
H_{jkl}=\kappa_G \,T_{jkl} \label{eq:Harada first}
\end{equation}
\begin{align*}
H_{jkl}=&\nabla_j R_{kl}+\nabla_k R_{lj}+\nabla_l R_{jk}\\
&-\tfrac{1}{3}(g_{kl}\nabla_j R+g_{lj}\nabla_k R+g_{jk}\nabla_l R)\\
T_{jkl}=&\nabla_j T_{kl}+\nabla_k T_{lj}+\nabla_l T_{jk}\\
&-\tfrac{1}{6}( g_{kl}\nabla_j T+g_{lj}\nabla_k T+g_{jk}\nabla_l T)
\end{align*}
$R_{jk}$ is the Ricci tensor with trace $R$, $T_{kl}$ is
the stress-energy tensor with trace $T$. 
The Bianchi identity $\nabla_{j}R^{j}{}_{k}=\frac{1}{2}\nabla_{k}R$
ensures $\nabla_{j}T^{j}{}_{k}=0$. Solutions of the Einstein equation
are solutions of the new theory. %In \cite{Harada23} the first vacuum solution was obtained.
%\subsection{The conformal Killing parametrization}

Shortly after their appearance, we found a parametrization showing that the Harada equations are equivalent to
the Einstein equations modified by a supplemental conformal Killing tensor $K_{kl}$ that is also divergence-free \cite{Mantica 23 a-1}: 
\begin{align}
 & R_{kl}-\tfrac{1}{2}Rg_{kl}= \kappa_G (T_{kl}+K_{kl})\label{eq:einstein enlarged}\\
 & \nabla_{j}K_{kl}+\nabla_{k}K_{jl}+\nabla_{l}K_{jk} \label{eq:Conformal Killing mantica-1}\\
 &=\tfrac{1}{6}(g_{kl}\nabla_{j}K+g_{jl}\nabla_{k}K+g_{jk}\nabla_{l}K). \nonumber
\end{align}
For this reason the theory was named Conformal Killing Gravity (CKG). The conserved tensor appears as a source term that is candidate for representing the yet unknown dark sector. 
%It modifies the TOV equation by the presence of a dark pressure and energy density.
%The reformulation makes the extension of GR explicit through the new term, that satisfies $\nabla^{k}K_{kl}=0$.
%For geometrical and physical applications of conformal Killing tensors see for example 
%\cite{Coll06,Kobialko22,Rani 03,Sharma10}.
%\begin{rem}

While $H_{jkl}$ in eq.\eqref{eq:Harada first} has third-order derivatives of the metric tensor, in the parametrization 
(\ref{eq:einstein enlarged}) %the Ricci tensor contains second order derivatives, while higher 
derivatives of order higher than two may appear in $K_{kl}$. 
In \cite{Mantica 23 a-1} we obtained a parametrization in Friedmann-Robertson-Walker background with a perfect fluid tensor $K_{kl}$ dependent
on the metric and not on derivatives. In that case the extended Friedmann equations %which include the conformal Killing divergence-free 
and the condition \eqref{eq:Conformal Killing mantica-1} were second order. They yielded the same forecast for the dark fluid obtained by Harada \cite{Harada23b}, and were tested with cosmological data \cite{Mantica 24}. 

So far, the majority of papers on CKG dealt with static spherical symmetric solutions of the third-order Harada equations. 
Barnes found solutions in vacuum and with Maxwell source \cite{Barnes23a,Barnes 24b}; Junior, Lobo and Rodriguez investigated regular black hole \cite{Junior24} and black 
bounce solutions \cite{Junior 24 b} in CKG coupled to nonlinear electrodynamics
and scalar fields. Vacuum cosmological solutions, wormholes and black holes were studied by Cl\'ement and Nouicer \cite{Clement24}.
Stability of thin-shell wormholes with CKG black holes was studied by Alshal et al. \cite{Alshal25}. \\
Differently, we found a conformal Killing parametrization in static spherically symmetry that keeps the field equations of second order \cite{Mantica 24b}. We 
%In particular we studied the density contrast, which behaves as $\Lambda$CDM
%in the matter dominated phase. The best fit for the CKG cosmological parameters with available data for the Hubble parameter %$H(z)$, predict future singularities. 
% with eqs.
%\eqref{eq:Harada first}
proved that the third order radial equations in the above quoted papers, are reducible to the second order extended Einstein equations.

The deduction of the CKG extension of the TOV equation and the study for a matter sphere in the dark energy ambient, are natural and simple in the conformal Killing parametrization, where the dark term associated  to the conformal Killing tensor is assimilated to an anisotropic fluid.\\ 

This is the plan of the paper.
In Sect.\ref{sec:Spherically-symmetric-static} we recollect the
covariant expressions of the Ricci tensor \cite{Mantica 23 b} and of the
conformal Killing tensor \cite{Mantica 24b} in static spherically symmetric
spacetimes, and write the field equations of CKG for anisotropic matter and dark fluid sources.\\
In Sect.\ref{sec:vacuum} we obtain the vacuum solutions by Cl\'ement and Nouicer, by solving a first-order differential equation, rather than a third-order one.
In Sect.\ref{sec:TOV equation} we obtain the extension of the TOV
equation in CKG sourced by an anisotropic fluid of ordinary matter. \\
%The original equation is modified by the contribution of the dark 
%energy density and pressure.\\
In Sect.\ref{STATICSPHERE} we write the equations for a static sphere of anisotropic fluid matter in CKG, and 
the junction conditions at the boundary of the matter sphere with the exterior CKG vacuum. They are obtained in the parametrization \eqref{eq:einstein enlarged}. Junction conditions for the original third order equations 
by Harada are probably obscure and difficult to achieve.
%The junction conditions in case of absence of thin matter shells or branes are also considered.\\
The equations are solved in Sect.\ref{UNIFORMSPHERE} for the sphere of perfect fluid with uniform matter density. 
The generalization of the Buchdahl limit for the mass of the critical star is obtained.\\
%Sect.\ref{NEWTON} we obtain the Newtonian limit of the static CKG equations for the uniform sphere and the vacuum.
%
%
%

\section{\label{sec:Spherically-symmetric-static} CKG equations in static\\ spherically symmetric 
spacetimes}
We quote an interesting extension of Birkhoff's theorem by Pavelle (1979) \cite{Pavelle79}: \textit{For a spherically symmetric space-time with  metric tensor 
$ ds^2 = - b^2(r,t)dt^2+ f_1^2(r,t)dr^2 + r^2 d\Omega^2_2 $, 
the necessary and sufficient conditions that Birkhoff's theorem is valid for the Einstein
equations $G_{ij} = \kappa_G T_{ij} $ are that the energy-momentum tensor is static and diagonal}.\\
In our parametrization \eqref{eq:einstein enlarged} of CKG, the conformal Killing tensor enters the Einstein equations as an additional source. Therefore, if $T_{ij}+K_{ij}$ is diagonal static and spherically symmetric, the metric is static. 
If  $T_{ij}=0$ (vacuum solution), then the CKG spherically-symmetric vacuum metric is static.

%With the choice $f_1(r)=1/b(r)$, the vacuum solution by Harada is \cite{Harada23}:
%\begin{align}
% b(r) = 1-\frac{2\overline M}{r} -\frac{\Lambda}{3} r^2 - \frac{\lambda}{5} r^4 \label{Hvacuum}
% \end{align}
%with constants $\overline M$, $\Lambda $ and $\lambda $.\\
%In \cite{Mantica 24b} eq.28 we showed that the most general solution is obtained for $f_1(r) = h(r)/b(r)$ where
%$h(r)=(\kappa_3 r^2 + \kappa_4)^{-1/2}$. This gives a parametric $b(r)$ that was first evaluated in \cite{Clement24} (eq. 2.17) via %Harada's third order equations, and by us, by solving the extended Einstein equation. 

We consider static spherically symmetric spacetimes with the following metric:
\begin{align}
ds^2=&-y(r) dt^2+\frac{h(r)}{y(r)} dr^2+r^2 d\Omega_2^2 \label{eq:Static sph symm}
%=& - b^2(r) dt^2 + f_1^2(r) dr^2 + r^2 d\Omega_2^2  \label{eq:Static sph symm_2}
\end{align}
Static spacetimes are covariantly characterized by the existence of a timelike unit vector $u_i$ (the velocity) and $\dot u_i=u^j\nabla_j u_i$ (the acceleration) with the properties \cite{Stephani}
$$ \nabla_i u_j = - u_i\dot u_j, \qquad \nabla_i \dot u_j = \nabla_j \dot u_i $$
In the (comoving) frame \eqref{eq:Static sph symm}: 
$u_{0}=-\sqrt y$, $u_{\mu}=0$. With the Christoffel symbols one evaluates $\dot u_0=0$, $\dot u_r= y'/(2y)$,
$\dot u_\theta=\dot u_\phi=0$: the acceleration is purely radial; it is normalized by
$\eta=\dot u^k\dot u_k=y'^2/(4hy)$. The unit vector $\chi_k =\dot u_k/\sqrt \eta$ is radial with nonzero component $\chi_r = \sqrt{h/y}$
in the comoving frame.
%A spherically symmetric (0,2) tensor is a combination of the tensors $u_i u_j$, $g_{ij}$ and $\dot u_i \dot u_j$.

The covariant expression of the Ricci tensor is eq. 85 in \cite{Mantica 23 b}:
\begin{align}
 R_{kl}=\frac{R+4\nabla_{p}\dot{u}^{p}}{3}u_{k}u_{l}+\frac{R+\nabla_{p}\dot{u}^{p}}{3}g_{kl} \label{eq:Ricci static}\\
 +\Sigma(r)\left[\chi_k \chi_l -\frac{u_{k}u_{l}+g_{kl}}{3}\right]. \nonumber
\end{align}
The curvature scalar $R$ of the space-time and $R^\star$ of the space-like submanifold, $\nabla_p \dot u^p$ and $\Sigma(r)$  
are respectively eqs. 90, 91, 87, 89 in \cite{Mantica 23 b} (here $f_1^2=h/y$ and $f_2^2=r^2$): 
\begin{align}
&R=R^{\star}-2\nabla_{p}\dot{u}^{p}\\
&\frac{R^\star}{2} = \frac{1}{r^2} + \frac{y}{r} \frac{h'}{h^2} - \frac{y'}{rh}- \frac{y}{r^2h} \label{EQRSTAR}\\
& \nabla_p \dot u^p = \frac{y''}{2h} - \frac{y'}{4} \frac{h'}{h^2} + \frac{y'}{rh} \label{EQNABLA}\\
& \Sigma = -\frac{y''}{2h} + \frac{y'}{4}\frac{h'}{h^2} + \frac{y}{r^2h} + \frac{y}{2r}\frac{h'}{h^2} - \frac{1}{r^2}  \label{EQSIGMA}
\end{align}
The following geometric identity results:
\begin{align}
\frac{R^\star}{2} +\nabla_p \dot u^p + \Sigma  = \frac{3y}{2r} \frac{h'}{h^2}  \label{GEOMETRIC}
\end{align}
%
%\subsection{Anisotropic Conformal Killing tensor }
In ref. \cite{Mantica 24b} we studied a symmetric tensor with the anisotropic structure of the Ricci tensor \eqref{eq:Ricci static}:
\begin{equation}
K_{kl}={\sf A}(r)u_{k}u_{l}+{\sf B}(r)g_{kl}+{\sf C}(r)\chi_k \chi_l \label{eq:anisotropic Conformal killing}
\end{equation}
%${\sf A}(r)$, ${\sf B}(r)$ and ${\sf C}(r)$ are named ``\textit{conformal Killing functions}''.\\
% $\chi_{j}=\dot u_j/\sqrt\eta$ is a unit space-like Vector, orthogonal to the velocity, $\chi^{p}u_{p}=0$. It is purely radial in the %comoving frame: $\chi_{0}=0$, $\chi_{r}=f_{1}$, $\chi_{\theta}=\chi_{\phi}=0$. \\
and proved a theorem that here reads:\\
{\bf Theorem:}\label{prop 5 Anisotropic} 
\textit{in the static spherically symmetric spacetime
 (\ref{eq:Static sph symm}) the tensor (\ref{eq:anisotropic Conformal killing})
is a divergence-free conformal Killing tensor if and only if 
\begin{equation}
\begin{array}{lr}
{\sf A}(r)=\kappa_{2}r^{2}-2\kappa_3 y (r)\\
\\
{\sf B}(r)=\kappa_1+ 2\kappa_2 r^2+\kappa_3 y(r) \\
\\
{\sf C}(r)= - \kappa_2 r^2
\end{array}\label{eq:funcion conformal killing}
\end{equation}
where $\kappa_1, \kappa_2, \kappa_3$ are constants (note the absence of $h(r)$ in the equations).}
%$A'+3B'+4C'=0$.

%\subsection{The conformal Killing gravity equations}
The structures of the Ricci tensor and of the conformal Killing tensor are compatible with an energy-momentum tensor for ordinary matter with the fluid form (see \cite{Mantica 24b}, Section 3)
\begin{align}
T_{kl}=&(\mu_m+P_m)u_k u_l + P_m g_{kl} \nonumber\\
&+(p_{mr}-p_{m\perp})\left[\chi_k \chi_l -\frac{g_{kl}+u_{k}u_{l}}{3}\right] \label{TMATT}
\end{align}
$\mu_m$ is the matter density, $p_{mr}$ and $p_{m\perp}$ are the radial and transverse pressures, $P_m=\frac{1}{3}(p_{mr}+2 p_{m\perp})$ is the
effective pressure. \\
In full analogy, the tensor $K_{kl}$ in \eqref{eq:anisotropic Conformal killing} may be rewritten as a ``dark''
anisotropic fluid:
\begin{align}
K_{kl}=& %(\mu_d+p_{d\perp})u_{k}u_{l}+p_{d\perp}g_{kl}+(p_{dr}-p_{d\perp})\frac{\dot u_k \dot u_l}{\eta}\label{eq:dark fluid}
(\mu_d+P_d)u_k u_l + P_d g_{kl} \nonumber\\
&+(p_{dr}-p_{d\perp})\left[\chi_k \chi_l -\frac{g_{kl}+u_{k}u_{l}}{3}\right]
\end{align}
with dark energy density $\mu_d$, radial and transverse dark pressures, effective dark pressure $P_d=\frac{1}{3}(p_{d r}+2p_{d \perp})$. They are related to the parameters of the metric tensor 
by the equations \eqref{eq:funcion conformal killing}:
\begin{equation}
\begin{array}{lr}
\mu_d={\sf A}-{\sf B}=-\kappa_1-\kappa_2 r^2 - 3\kappa_3 y(r)\\
\\
p_{d r}={\sf B}+{\sf C}=\kappa_1+\kappa_2 r^2 + \kappa_3 y(r)\\
\\
p_{d \perp}={\sf B}=\kappa_1+ 2\kappa_2 r^2+\kappa_3 y(r)
\end{array}\label{eq:dark parameters}
\end{equation}
%
%=\frac{5}{3}\kappa_{2}f_{2}^{2}+\kappa_{3}b^{2}+\kappa_{1}$.\\ 
The dark energy density can be negative. 
Note the identity (the ``dark equation of state''):
\begin{align}
 \mu_d(r) +5 p_{dr} (r) - 2 p_{d\perp} (r) =  2\kappa_1. \label{DEOS}
 \end{align}
The field equations  of CKG ($\kappa_G$ is included in
the fluid parameters) $R_{kl}-\tfrac{1}{2}Rg_{kl}= T_{kl}+K_{kl}$ provide three scalar equations: 
\begin{equation}
\begin{array}{lr}
\frac{1}{2}R^{\star}=\mu_m+\mu_d\\
\\
\nabla_p \dot u^p = \frac{3}{2}(P_m+P_d)+\frac{1}{2}(\mu_m+\mu_d)\\
\\
\Sigma=(p_{mr}-p_{m\perp})+(p_{d r}-p_{d \perp})
\end{array}\label{eq:General conformal Killing equations-1}
\end{equation}
Note that $\frac{1}{2}R^\star +\nabla_p \dot u^p + \Sigma = \frac{3}{2}(\mu_m + \mu_d + p_{mr} + p_{dr})$, and $\mu_d +p_{dr} = -2\kappa_3 r$. Using the geometric identity \eqref{GEOMETRIC} we obtain:
\begin{align}
\mu_m + p_{mr} = \frac{y(r)}{r} \frac{d}{dr} \left[ \kappa_3 r^2 -\frac{1}{h(r)} \right ]. \label{EQFORh}
\end{align}
The matter conservation law $\nabla_k T^{jk}=0$ in static radial symmetry is the scalar equation $\dot u_j \nabla_k T^{jk}=0$. With the 
expression \eqref{TMATT} one obtains the equation for the derivative of the radial pressure:
\begin{align}
\frac{dp_{mr}}{dr} = - \frac{y'}{2y} (\mu_m + p_{mr} ) -   \frac{2}{r} (p_{mr} - p_{m\perp}). \label{CONTINUITY}
\end{align}
In a perfect fluid  the pressures are equal: $p_{mr}=p_{m\perp}\equiv p_m$.

\section{\label{sec:vacuum} The vacuum solutions revisited}
In absence of ordinary matter, the field equations \eqref{eq:General conformal Killing equations-1} become the vacuum CKG static spherical  equations:
\begin{align}
&\tfrac{1}{2}R^\star = - \kappa_1 -\kappa_2 r^2 -3 \kappa_3 y(r)\\
& \nabla_p \dot u^p = \kappa_1 + 2\kappa_2 r^2\\
& \Sigma = -\kappa_2 r^2 \label{EQ3}
\end{align} 
$\kappa_1$, $\kappa_2$, $\kappa_3$ are arbitrary constants; they are zero in GR.\\
Equation \eqref{EQFORh} gives:
\begin{align}
h(r)  = \frac{1}{\kappa_3 r^2 + \kappa_4} 
\end{align}
where $\kappa_4$ is a new constant. In GR $\kappa_3=0$ and $\kappa_4=1$. In CKG if $\kappa_3 >0$
one may choose $\kappa_4=0$.

By entering the expressions for $R^\star$, $\nabla_p\dot u^p$ and $\Sigma$, 
the CKG vacuum field equations become:
\begin{align}
&  \frac{y'}{r} (\kappa_3 r^2 +\kappa_4) + \frac{y}{r^2}\kappa_4 = \kappa_1 +\kappa_2 r^2 + \frac{1}{r^2} \label{EQUATION1}\\
&y''(\kappa_3 r^2 +\kappa_4) + \frac{y'}{r} (3\kappa_3 r^2 +2 \kappa_4) = 4 \kappa_2 r^2 
+ 2\kappa_1 \label{EQUATION2}\\
&y''(\kappa_3 r^2 + \kappa_4)  + y'\kappa_3 r  - 2 \frac{y}{r^2}\kappa_4 = 2\kappa_2 r^2  -  \frac{2}{r^2} \label{EQUATION3}
\end{align} 
The equations are not independent: subtraction of the third equation from the second one gives the first equation. We then drop the second equation. The derivative of
the first equation gives the third equation. Therefore only the first equation has to be solved, and it is of first order!

We begin with two special cases:\\
$\kappa_4=1$, $\kappa_3=0$.\\ 
Eq.\eqref{EQUATION1} is: $\frac{y'}{r}  +  \frac{y}{r^2} = \kappa_1 + \kappa_2r^2 +  \frac{1}{r^2}$. This case was solved by Harada  (eq.21 in \cite{Harada23}, with  $C=-2\overline M$, $\kappa_1 =-\Lambda $, $\kappa_2=-\lambda $):
\begin{align}
 y(r)= 1-\frac{2\overline M}{r} - \frac{\Lambda}{3} r^2 - \frac{\lambda}{5} r^4  \label{HARADAVAC}
\end{align}
It is the Schwarzschild solution for $\lambda=\Lambda=0$, $\overline M>0$. The positive roots of $y(r)=0$ define one or more horizons that bound the regions where $y(r)>0$, studied in \cite{Alshal25} .\\
$\kappa_4=0$, $\kappa_3>0$.\\
Eq.\eqref{EQUATION1} is $ y'    =\frac{ \kappa_1}{\kappa_3} \frac{1}{r} +\frac{\kappa_2}{\kappa_3} r +\frac{1}{\kappa_3} \frac{1}{r^3} $ with solution
\begin{align} 
y(r) = \frac{ \kappa_1}{\kappa_3} \log (r \sqrt{\kappa_1}) +\frac{\kappa_2}{2\kappa_3} r^2 - \frac{1}{2\kappa_3} \frac{1}{r^2}  +{\rm const.} \label{SPECIAL}
\end{align}
It is positive for $r$ beyond an horizon if $\kappa_1>0$ and $\kappa_2>0$. For $\kappa_2=0$ it grows as $\log r$ and
the solution can be useful in the problem 
of galactic rotation curves.

The solutions for $\kappa_3\kappa_4\neq 0$, $\kappa_3r^2+\kappa_4\ge 0$, were obtained by Cl\'ement and Noucier \cite{Clement24}. They solved a third order Harada radial equation. We obtain them by solving  the simpler first order equation \eqref{EQUATION1}. The calculations are in Appendix 1.

%\begin{prop}\label{GENSOLUTION}\quad\\
\begin{align}
y(r)&= C \frac{\sqrt{\kappa_3 r^2+\kappa_4}}{r}  +\frac{1}{\kappa_4}+ \frac{\kappa_2}{2\kappa_3} r^2 \label{GENERAL}\\
& +\left[ \frac{\kappa_1}{\kappa_3} - \frac{3}{2} \frac{\kappa_2\kappa_4}
{\kappa_3^2} \right ]
\left [  \frac{\sqrt{\kappa_3 r^2+\kappa_4}}{r\sqrt{|\kappa_3 |}} F\left (r \sqrt{ \frac{|\kappa_3|}{|\kappa_4|}} \right ) -1\right ]  
\nonumber
\end{align}
$$
F(x) =\begin{cases} {\rm ArcSinh} \,x&  \kappa_3>0, \kappa_4>0 \\
{\rm ArcCosh} \,x &  \kappa_3>0, \kappa_4<0 \\
{\rm ArcSin}\,x &  \kappa_3<0, \kappa_4>0 \\
\end{cases} \nonumber
$$
%\end{prop}
The behaviour of the solutions with sh and ch for $r^2\gg\frac{ |\kappa_4|}{\kappa_3}$ is:
\begin{align*}
 y(r) \approx &\frac{\kappa_2}{2\kappa_3} r^2 + \left[ \frac{\kappa_1}{\kappa_3} - \frac{3}{2} \frac{\kappa_2\kappa_4}
{\kappa_3^2} \right ] \log \left ( 2r \sqrt{ \frac{\kappa_3}{|\kappa_4|}} \right )\\
  &+ \left[ \frac{1}{\kappa_4} - \frac{\kappa_1}{\kappa_3} + \frac{3}{2} \frac{\kappa_2\kappa_4}{\kappa_3^2} \right ]  + ...
  \end{align*}
The special cases \eqref{HARADAVAC}, \eqref{SPECIAL} result from the sh solution in the limits $\kappa_3\to 0$, $\kappa_4=1$ or $\kappa_3>0$, $\kappa_4\to 0$.\\

\begin{figure}[h]
\begin{center}
\includegraphics*[width=6cm,clip=]{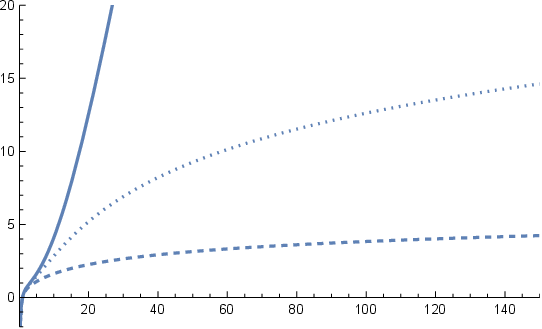}
\includegraphics*[width=6cm,clip=]{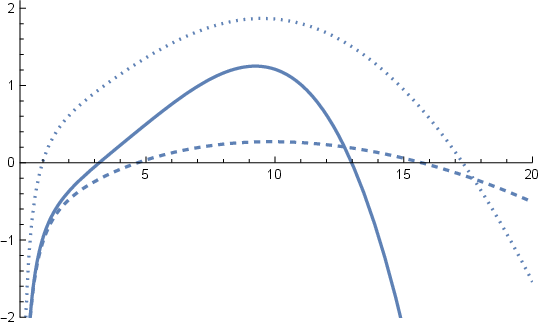}
\caption{The function $y(r)$ with ArcSinh with $C=-1$, $\kappa_1=0.1$, $\kappa_2=0$ (left) and $\kappa_2=-0.001$ (right), $\kappa_4=1$, for $\kappa_3=\frac{1}{1000}$ (full), $\kappa_3=\frac{1}{50}$ (dotted), 
$\kappa_3=\frac{1}{10}$ (dashed). For small $r$ the functions diverge as $-1/r$. In left panel ($\kappa_2=0$), for large $r$ they grow as logarithms; the first case mimics Harada's solution with apparent quadratic growth. There is a single horizon near the origin. In the right panel ($\kappa_2<0)$, there is a second horizon where $y$ turns negative.}
\end{center}
\end{figure}

%The solution can be reformulated in the adimensional variable $x={\rm sh}\theta = r/r_0$:
%\begin{align*}
% y(r) =  (C  \sqrt{\kappa_3}) \frac{\sqrt{1+x^2}}{x} +
% \left(\frac{\kappa_1}{\kappa_3} -\frac{3}{2}\frac{\kappa_2\kappa_4}{\kappa_3^2} \right)\left[\frac{\sqrt{1+x^2}}{x}{\rm Arsh}x-1\right]+ 
% \frac{1}{\kappa_4}+\frac{\kappa_2\kappa_4}{2\kappa_3^2} x^2
% \end{align*}

\section{\label{sec:TOV equation}The Tolman-Oppenheimer-Volkoff equation
for CKG }

We obtain the extension of the TOV equation in conformal Killing gravity sourced by ordinary matter described by the anisotropic tensor \eqref{TMATT}.\\
We parametrize the static spherically symmetric metric as follows:
\begin{equation}
ds^{2}=-b^{2}(r)dt^{2}+\left[1-\frac{2M(r)}{r}\right]^{-1}dr^{2}+r^{2}d\Omega_{2}^{2}\label{eq:metric phi-1}
\end{equation}
The expressions \eqref{EQRSTAR} for $R^\star$,  \eqref{EQNABLA} for $\nabla_p \dot u^p$, and \eqref{EQSIGMA} for $\Sigma$,
are evaluated with $y=b^2$ and $y/h =  1-2M(r)/r$:
\begin{align}
& R^\star = \frac{4M'}{r^2} \label{eq:spacial-1} \\
& \Sigma = -\left [ \frac{b''}{b}-\frac{b'}{br} \right] \left [1-\frac{2M(r)}{r} \right] + \frac{b'}{b} \left[\frac{M' r- M}{r^2}\right ]
 \nonumber\\
&\qquad -\frac{3M-rM'}{r^3} \label{eq:anisotropic-1}\\
& \nabla_p \dot u^p = \left[ \frac{b''}{b} + \frac{2b'}{br} \right] \left [ 1- \frac{2M(r)}{r} \right] - \frac{b'}{b}
\left[\frac{M'r-M}{r^2} \right]\label{eq:div dot u}
\end{align}
The first equation is integrated with the first one in \eqref{eq:General conformal Killing equations-1} and the assumption  $M(0)=0$:
\begin{align}
M(r) = \frac{1}{2} \int^r_0 dr'  r'^2 [\mu_m (r')+ \mu_d (r')] \label{Mr}
\end{align}

%\subsection{Properties of metric function%
%We are in position to obtain the extended Tolman-Oppenheimer-Volkoff (TOV) equation.
{\bf Theorem:}\label{Extended TOV} 
\textit{in a static spherically symmetric spacetime sourced by an anisotropic matter fluid, the following  conformal Killing gravity 
extension of the Tolman-Oppenheimer-Volkoff equation holds:
\begin{align}
&\left [ p_{m r}'  +  2 \frac{p_{mr} - p_{m\perp}}{r}\right ]
\left[1-\frac{2M(r)}{r}\right ]\nonumber \\
& = -\frac{\mu_m + p_{mr}}{2} 
 \left[ r (p_m+p_{d r})+\frac{2M(r)}{r^2} \right] 
  \label{ExtendedTOV}
\end{align}
where $M(r)$ is eq.\eqref{Mr}.}\\
\textit{Proof.}
The expressions for $R^\star$, $\Sigma$ and $\nabla_p \dot u^p$ imply the identity: 
\begin{equation}
\nabla_p \dot u^p+\Sigma-\frac{R^\star}{4} = \frac{3b'}{br} \left[1-\frac{2M(r)}{r} \right] -\frac{3M(r)}{r^3} \label{eq:geometric result}
\end{equation}
By the field equations \eqref{eq:General conformal Killing equations-1} it is also $\nabla_p \dot u^p + \Sigma - \frac{R^\star}{4}=\frac{3}{2} (p_{mr}+p_{d r})$. 
This and the continuity equation \eqref{CONTINUITY} for $b'/b=y'/2y$, are used in \eqref{eq:geometric result} to obtain the extension of the 
TOV equation \eqref{ExtendedTOV}. \hfill $\square $

%\noindent
If $\mu_d=p_{dr}=0$ the anisotropic TOV equation for GR is recovered (eq.53 in \cite{Herrera13}). The isotropic TOV equation \eqref{TOV_GR}  (in units $\kappa_G=1$, $c=1$) is obtained for $p_{mr}=p_{m\perp}=p_m$.
%\[ p'\left [1-\frac{2M(r)}{r}\right ]=-\frac{p+\mu}{2}\left [rp+\frac{2M(r)}{r^2}\right ]. \]

%\section{\label{sec: continuity conformal Killing tensor} Discontinuity of the CKT and junction conditions}
\section{Static sphere in CKG  gravity:\\ the junction conditions}\label{STATICSPHERE}
We apply the field equations \eqref{eq:General conformal Killing equations-1} of static spherical CKG to study a sphere of radius $R$ of ordinary matter (a star), modelled by the anisotropic tensor \eqref{TMATT} with exterior given by a static spherical vacuum solution of CKG. \\
The interior and exterior metrics are:
\begin{align}
& ds_-^{2}=-b_-^{2}(r)dt^{2}+\frac{dr^2}{1-\dfrac{2M(r)}{r}}+r^2 d\Omega_2^2 & r<R\label{interior}\\
%& ds_+^{2}=-b_+^{2}(r)dt^{2}+\frac{dr^2}{b_+^2(r)}+r^2 d\Omega_2^2 & r> R \label{exterior}
& ds_+^{2}=-b_+^{2}(r)dt^{2}+f_{1+}^2 (r) dr^2 +r^2 d\Omega_2^2 & r> R \label{exterior}
\end{align}
%where $b_+^2(r)$ is the Harada solution eq.\eqref{Hvacuum}, with parameters $M$, $\lambda$, $\Lambda$.\\
The metric functions satisfy junction conditions at $r=R$: continuity of the metric and continuity of the exterior derivative.
The latter becomes a balance condition for the radial pressures.

\begin{prop}[The junction conditions]
\begin{gather}
b_-^2(R) =  b^2_+(R)\label{COND0}\\
1-\dfrac{2M(R)}{R} = \frac{1}{f_{1+}^2(R)}  \label{COND1}\\
% b_-^2(R) = 1-\dfrac{2M(R)}{R} = 1-\frac{2\overline M}{R} - \frac{\Lambda}{3}R^2 -\frac{\lambda}{5} R^4  \label{COND1}\\
 p_{mr}(R) = p_{dr}^+ (R) - p_{dr}^-(R) \label{COND2}
\end{gather}
\end{prop}
\textit{Proof.}
The first two conditions  are the continuity of the metric %$b_+(R)=b_-(R)$ and $f_{1+}(R)=f_{1,-}(R)$ 
(see for example eq.~3 in \cite{Santos 85}, or \cite{Senovilla 13}).\\
The third condition requires some steps.

In the comoving frame the velocity has only the time component: $u_0^\pm(r)=-b_\pm (r)$. Continuity of the metric at $r=R$ implies continuity of the four-vectors: 
$u_{k}^{-}(R)=u_{k}^{+}(R)\equiv u_{k}(R)$. Similarly, the normalized acceleration has only the radial component: 
$\chi_r^\pm (r)=f_1^\pm (r) $; continuity implies $\chi_{k}^{-}(R)=\chi_{k}^{+}(R)\equiv \chi_{k}(R)$.\\
In principle the interior and the exterior conformal Killing tensors are different:
\begin{align*}
K_{kl}^{\pm}(r)=&{\sf A}^{\pm}(r)u_{k}^{\pm}(r)u_{l}^{\pm}(r)\\
&+{\sf B}^{\pm}(r)g_{kl}^{\pm}(r)+{\sf C}^{\pm}(r)\chi_{k}^{\pm}(r)
\chi_{l}^{\pm}(r)
\end{align*}
The jump at the boundary surface is:
\begin{align}
&\delta K_{kl}(R) = K_{kl}^{+}(R)-K_{kl}^{-}(R) \label{eq:jump for CKT}\\
&=\delta {\sf A}(R) u_{k}(R)u_{l}(R)+\delta {\sf B}(R) g_{kl}(R)
+\delta {\sf C}(R) \chi_{k}(R)\chi_{l}(R)\nonumber
\end{align}
where 
\begin{equation}
\begin{array}{lr}
\delta {\sf A}(R) = %{\sf A}^{+}(R)-{\sf A}^{-}(R)=
(\kappa_{2}^{+}-\kappa_{2}^{-})R^2 -2(\kappa_{3}^{+}-\kappa_{3}^{-})b^{2}(R)\\
\\{}
\delta {\sf B}(R) = %{\sf B}^{+}(R)-{\sf B}^{-}(R)=
(\kappa_{1}^{+}-\kappa_{1}^{-})+2(\kappa_{2}^{+}-\kappa_{2}^{-})R^2+(\kappa_{3}^{+}-\kappa_{3}^{-})b^{2}(R)\\
\\{}
\delta {\sf C}(R) = %{\sf C}^{+}(R)-{\sf C}^{-}(R)=
-(\kappa_{2}^{+}-\kappa_{2}^{-})R^2
\end{array}\label{eq:jump for CKT 2}
\end{equation}
and $b^2(R)=b^2_\pm (R)$.
%\begin{prop}
%The conformal Killing tensor at the surface $r=R$ exhibits at most
%a jump discontinuity as in (\ref{eq:jump for CKT}) and (\ref{eq:jump for CKT 2})
%\end{prop}

%Remember now that in absence of matter shells or branes, 
The continuity of the second fundamental form (extrinsic curvature) at $r=R$ is  (see \cite{Senovilla 13}, Appendix 3, or \cite{Gabbanelli 19} eq. 2.31):
$$\nabla^-_k\chi^-_l (R) = \nabla^+_k \chi^+_l (R) $$
%
%\begin{equation}
%\delta(\nabla_k \chi_l)(R)=0\label{eq:   continuitysecond fund form}
%\end{equation}
% is equivalent to the vanishing of the whole singular part of the Riemann tensor; moreover the same condition 
It implies that the jump discontinuity of the Riemann tensor has the form (eq.~A20 in \cite{Senovilla 13})
\[
\delta R_{jklm} =\chi_j \chi_l B_{km} - \chi_j \chi_m B_{kl} + \chi_k \chi_m B_{jl} - \chi_k \chi_l B_{jm}
\]
where $B_{km}$ is a symmetric tensor defined at the boundary such that $\chi^{k}B_{km}=0$.\\
The discontinuity for the Ricci tensor is $\delta R_{kl}=B_{kl}+\chi_k \chi_l B$,
being $B=g^{kl}B_{kl}$, and  $\delta R=2B$. Finally,
the discontinuity for the Einstein tensor is $\delta G_{kl}=B_{kl}+B(g_{kl}-\chi_k \chi_l).$
Thus it is inferred that (\cite{Senovilla 13} eq. A21 or \cite{Santos 85}
eq. 6, or \cite{Gabbanelli 19} eq. 2.32):
\begin{equation}
0=\delta G_{kl}(R) \chi^{k}=G_{kl}^{+}(R)\chi^{k}-G_{kl}^{-}(R)\chi^{k}
\end{equation}
In view of the conformal Killing equations (\ref{eq:einstein enlarged})
$G_{kl}^{-}=T_{kl}+K_{kl}^{-}$ and $G_{kl}^{+}=K_{kl}^{+}$, 
%Moreover %from eq 12 in \cite{Alshal 24} it is $n_{k}=(0,f_{1}(r),$0,0)
%we make the identification $n_{k}=\chi_{k}$ as a unit space-like vector orthogonal to the separation surface. Thus 
we get the junction
condition $T_{kl} \chi^k (R) = \delta K_{kl}(R)\chi^k(R)$ i.e.
\begin{align*}
p_{mr} (R)  &= \delta {\sf B} (R)+ \delta {\sf C} (R)\\
&=(\kappa_1^+-\kappa_1^-)+(\kappa_2^+-\kappa_2^- ) R^2 + (\kappa_3^+-\kappa_3^-)b^2(R)
\end{align*}
Using the second of (\ref{eq:dark parameters}) one obtains eq.\eqref{COND2}. \hfill $\square$

In GR the external metric is Schwarzschild, with $b^2_+(r) =1/f_{1+}^2(r) = 1-2\overline M/r$. The junction conditions reduce to $\overline M = M(R)$ and $p_{mr}(R)=0$.  

\begin{prop}
If the conformal Killing tensor is continuous at the surface $r=R$, then: 
\begin{align*}
&\kappa_{1}^{+}=\kappa_{1}^{-},\quad \kappa_{2}^{+}=\kappa_{2}^{-}\quad \kappa_{3}^{+}=\kappa_{3}^{-}\\
&p_{mr}(R)=0.
\end{align*}
\end{prop}
\textit{Proof.}
If the conformal Killing tensor is continuous at $r=R$, then \eqref{eq:jump for CKT} gives
\begin{equation*}
\delta{\sf A}(R)u_{k}(R)u_{l}(R)+\delta{\sf B}(R)g_{kl}(R)+\delta {\sf C}(R)\chi_{k}(R)\chi_{l}(R)=0
\end{equation*}
The $\theta\theta$ component of this relation gives $\delta{\sf B}(R)=0$;
the $tt$ component gives $\delta {\sf A}(R) b^{2}(R)-\delta {\sf B}(R)b^{2}(R)=0$
from which $\delta {\sf A}(R)=0$. Finally the $rr$ component gives
$\delta {\sf B}(R) f_{1}^{2}(R)+\delta {\sf C}(R)f_{1}^{2}(R)=0$ from
which $\delta {\sf C}(R)=0$. This last condition implies $\kappa_{2}^{+}=\kappa_{2}^{-}$.
This is inserted in $\delta {\sf A}(R)=0$ to infer $\kappa_{3}^{+}=\kappa_{3}^{-}$;
finally from $\delta {\sf B}(R)=0$ it is $\kappa_{1}^{+}=\kappa_{1}^{-}$. 
It follows that $p_{mr}(R)=0$. \hfill $\square $ 

\section{Perfect fluid star\\ in Harada vacuum}\label{UNIFORMSPHERE}

Hereafter, the exterior solution is the Harada vacuum \eqref{HARADAVAC}. Continuity of the metric eq.\eqref{COND1} gives
 \begin{align}
 1-\dfrac{2M(R)}{R} = 1-\frac{2\overline M}{R} - \frac{\Lambda}{3}R^2 -\frac{\lambda}{5} R^4  \label{CONDH}
\end{align}
The star with radius $R$ is a perfect fluid matter with mass density $\mu_m(r)$ and pressure $p_m(r)$.
The mass integral \eqref{Mr} at $R$ and the junction condition \eqref{COND1}  give:
\begin{align}
\frac{1}{2}\int_0^R dr' [\mu_m(r')+\mu_d(r')]r'^2  =  \overline M + \frac{\Lambda}{6} R^3 + \frac{\lambda}{10} R^5 
\label{continuity}
\end{align}
{\em Let us suppose that the conformal Killing tensor is continuous at the surface} $r=R$. Then $p_{mr}(R)=0$ 
and the inner and outer constants are equal: $\kappa_j^+=\kappa_j^-\equiv \kappa_j$ ($j=1,2,3$).\\
As shown in \cite{Mantica 24b} Example 2, the form of the Harada vacuum metric corresponds to the 
following specification of  $K_{kl}^+$: 
$$  \kappa_{1}=-\Lambda, \quad  \kappa_{2}=-\lambda, \quad \kappa_3=0 $$ 
By continuity, these values hold also in the inner region. Therefore, the dark fluid density and pressures are the same functions inside and outside the star:
\begin{align}
&\mu_d(r) = \lambda r^2 +\Lambda\\
& p_{dr}(r)=-\lambda r^2 -\Lambda\\
& p_{d\perp}(r) = -2\lambda r^2 -\Lambda
\end{align}
Eq.\eqref{continuity} simplifies and shows that $\overline M$ is the total matter mass:
\begin{align}
\frac{1}{2}  \int_0^R dr \, r^2 \mu_m(r)   =  \overline M 
\end{align}
\\ 
The other internal field equations are: $\Sigma(r) = p_{dr}-p_{d\perp} = \lambda r^2$, i.e.
\begin{align}
\left [1-\frac{2M(r)}{r}\right ] b_-'' - \left [\frac{1}{r} + \frac{rM'(r)-3M(r)}{r^2}\right ] b_-'  \nonumber\\
+\left [\lambda r^2 -  \frac{rM'(r)-3M(r)}{r^3}\right ] b_- =0 \label{SIGMA}
\end{align}
and $\nabla_p \dot u^p = \frac{3}{2}p_m +\frac{1}{2} \mu_m - 2\lambda r^2 -\Lambda $ i.e.
\begin{align}
\left[1-\frac{2M(r)}{r}\right]b''_- + \left[\frac{2}{r} - \frac{rM'(r)+3M(r)}{r^{2}}\right]  b'_- \nonumber \\ 
 - \left[\frac{3}{2}p_m +\frac{1}{2} \mu_m - 2\lambda r^2 -\Lambda\right ] b_-=0 \label{NABLA}
\end{align}
%
%$T_{kl}=(\mu_m+p_m)u_k u_l+p_mg_{kl}$. \\
% The field equations \eqref{eq:General conformal Killing equations-1} simplify
%\begin{equation}
%\begin{array}{lr}
%\frac{1}{2}R^{\star}=\mu_m+\mu_d\\
%\\
%\nabla_{p}\dot{u}^{p}=\frac{3}{2}(p_m + P_d)+\frac{1}{2}(\mu_m+\mu_d)\\
%\\
%\Sigma=p_{d r}-p_{d \perp}
%\end{array}\label{eq:conformal Killing perfect fluid}
%\end{equation}
The continuity equation \eqref{CONTINUITY} for the perfect fluid it is:
\begin{equation}
\frac{p_m'}{p_m+\mu_m}=-\frac{b'}{b}\label{eq:cont eq}
\end{equation}

%\section{Star with uniform matter density in CKG}\label{UNIFORMSPHERE}
To proceed, we consider a {\em star with  uniform mass density $\mu_m$}. Then:
$$M(r)=\frac{\mu_m +\Lambda}{6} r^3 +\frac{\lambda}{10}r^5$$
Note that the same polynomial expression of  $M(r)$, with negative $\lambda $, appears as eq.50 in \cite{Herrera24} in the study of ``ghost stars''. They have total mass $M(R)=0$ by the presence of regions of negative energy density in the fluid distribution.

Equation \eqref{eq:cont eq} is straightforwardly integrated and links the
inner function $b_-(r)$ with the matter pressure, through a constant $\gamma$:  
\begin{align}
b_-(r) = \gamma \frac{\mu_m}{p_m(r) + \mu_m} \label{bminus}
\end{align}
The boundary conditions $b_-(R)=b_+(R)$ and $p_m(R)=0$ give:
\begin{align}
\gamma =\sqrt{1-\frac{2M(R)}{R} }= \sqrt{1 -\frac{\mu_m+\Lambda}{3}R^2 - \frac{\lambda}{5}R^4 }
\end{align}
%and $\overline M = \frac{1}{6}\mu_m R^3$. \\
The TOV equation (\ref{ExtendedTOV}), with $p_{d r}=-\lambda r^{2}-\Lambda$, is a Bernoulli equation for  $y(r)=p_m(r)+\mu_m$:
\begin{equation}
y'\left[1-\frac{\mu_m+\Lambda}{3}r^2-\frac{\lambda}{5}r^4 \right ]+ \frac{y^2}{2} r - y\left[\dfrac{\mu_m+\Lambda}{3} r+2\frac{\lambda}{5}r^3\right] =0  \label{TOVUNIF}
\end{equation} 
The solution (see Appendix 2) gives the internal matter pressure, where the junction condition $p_m(R)=0$ is used to fix a constant:
\begin{align} 
\frac{1}{p_m(r)+\mu_m} &=  \frac{1}{4}  \frac{\frac{\mu_m+\Lambda}{6}+\frac{\lambda}{5}r^2}{(\frac{\mu_m+\Lambda}{6})^2+\frac{\lambda}{5}}  \label{resultTOV}\\
&+ \sqrt{ \frac{1-\frac{2M(r)}{r} }{ 1 - \frac{2M(R)}{R}}} 
 \left [ \frac{1}{\mu_m} - \frac{1}{4}  \frac{\frac{\mu_m+\Lambda}{6}+\frac{\lambda}{5}R^2}{(\frac{\mu_m+\Lambda}{6})^2+\frac{\lambda}{5} } \right ] \nonumber
\end{align}
%
%For $\Lambda \ge 0$ and $\lambda \ge 0$ the function is positive in $r\le R$.\\
For $\Lambda=0$, $\lambda =0$ one obtains the GR relation, with $M(r)= \frac{\mu_m}{6}r^3$:
\begin{align*} 
p_m(r) = \mu_m \frac{ \sqrt{1-2M(R)/R} -  \sqrt{1-2M(r)/r}}{ \sqrt{1-2M(r)/r} - 3 \sqrt{1-2M(R)/R}}
\end{align*}
The central pressure is divergent at the Buchdahl limit $M(R)=\frac{4}{9}R$ \cite{Buchdahl59}.

%For $r=0$ one reads the central pressure:
%\begin{align} 
%\frac{1}{p_m(0)+\mu_m} =
% \frac{1 }{ \sqrt{ 1 - \frac{2M(R)}{R}}} 
% \left [ \frac{1}{\mu_m}+\frac{1}{4} \frac{\frac{\mu_m+\Lambda}{6}+\frac{\lambda}{5}R^2}{(\frac{\mu_m+\Lambda}{6})^2+
%\frac{\lambda}{5} } \right ] -  \frac{1}{4} \frac{\frac{\mu_m+\Lambda}{6}}{(\frac{\mu_m+\Lambda}{6})^2+\frac{\lambda}{5}} 
%\end{align}
%
By eq.\eqref{bminus} the pressure fixes the interior metric in terms of the parameters $\mu_m$, $R$ and the parameters 
$\lambda $ and $\Lambda$
of the outer metric:
\begin{align}
b_-(r) =  
 \sqrt{ 1 - \frac{2M(r)}{r}} 
 \left [1-\frac{\mu_m}{4}  \frac{\frac{\mu_m+\Lambda}{6}+\frac{\lambda}{5}R^2}{(\frac{\mu_m+\Lambda}{6})^2+\frac{\lambda}{5} } \right ]\nonumber\\ 
 + \frac{\mu_m}{4}  \frac{\frac{\mu_m+\Lambda}{6}+\frac{\lambda}{5}r^2}{(\frac{\mu_m+\Lambda}{6})^2+\frac{\lambda}{5}}  \sqrt{ 1 - \frac{2M(R)}{R}} 
\end{align}
$b_-(r)$ is a linear combination of the radial functions $\sqrt{1-2M(r)/r}$ and $\frac{\mu_m+\Lambda}{6}+\frac{\lambda}{5}r^2$. The two functions are independent solutions of eq.\eqref{SIGMA}:
\begin{align*}
\left [1-\frac{\Lambda+\mu_m}{3}r^2 -\frac{\lambda}{5} r^4\right ] b_-'' - \left [\frac{1}{r} + \frac{\lambda}{5}r^3\right ] b_-' + \frac{4\lambda}{5} r^2  b_- =0 
\end{align*}
Subtraction of eq.\eqref{SIGMA} from \eqref{NABLA} and use of $b(p_m+\mu_m)=\gamma \mu_m$ gives the equation
$$\left [1-\frac{2M(r)}{r}\right  ] b'_- +2 \left [\frac{\mu_m+\Lambda}{6} r 
+ \frac{\lambda}{5} r^2\right ] b_- = \frac{r}{2}\gamma \mu_m$$
which coincides with the TOV equation \eqref{TOVUNIF} divided by $y^2$, which becomes an equation for $1/y(r)$ and
$b_-(r) = \gamma \mu_m /y(r)$. This proves consistency of the solution $b_-$ with the field equations.

The solution of the uniform sphere is fixed by the free parameters of the star (density $\mu_m$ and radius $R$), 
and the parameters $\Lambda$ and $\lambda $ of the dark fluid, which fills with
continuity the matter sphere and outer space.\\
The outer metric, besides the dark parameters that deviate it from the Schwarzschild metric, has the mass parameter 
$\overline M = \frac{1}{2} \int_0^R dr  r^2 \mu_m = \frac{1}{6}\mu_m R^3$.

\begin{figure}[t]
\begin{center}
\includegraphics*[width=4cm,clip=]{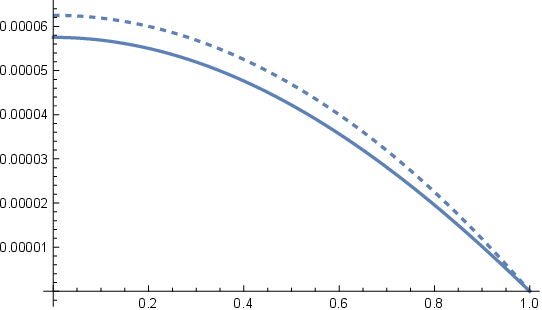}\quad
\includegraphics*[width=4cm,clip=]{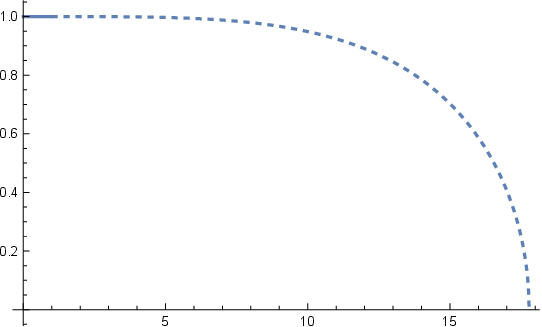}\\
\includegraphics*[width=4cm,clip=]{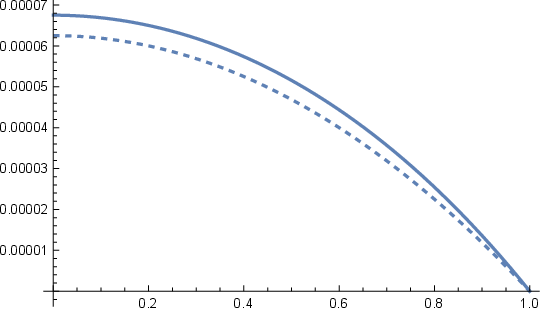}\quad
\includegraphics*[width=4cm,clip=]{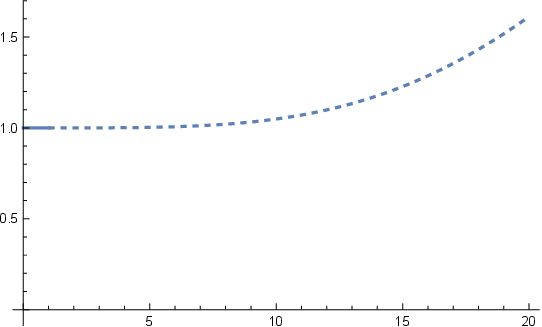}
\caption{\label{PRESS} Left: the pressure $p_m(r)/\mu_m$ in the homogeneous sphere in the Harada dark fluid (full) and in Schwarzschild vacuum (dashed), as functions of $r/R$. Right: the inner metric function $b_-$ (full) and the Harada vacuum $b_+$ (dashed) as functions of $r/R$. They join at $r/R=1$, at an angle not visible 
in the picture. 
Upper pictures: $\Lambda =0$,  $X=\frac{1}{3}\mu_m R^2 =\frac{1}{4}10^{-3}$, $Y=\frac{1}{5}\lambda R^4= 10^{-5}$,  (in GR: $X=\frac{1}{4} 10^{-3}$, $Y=0$). Here the Harada metric becomes negative around $r/R= 18$. 
Lower pictures: $\Lambda =0$,  $X=\frac{1}{4}10^{-3}$, $Y=- 10^{-5}$. }
\end{center}
\end{figure}

The pressure functions in GR or CKG of the homogeneous star in the Schwarzschild or Harada vacuum $y(r)=1-\frac{2\bar M}{r} -\frac{\lambda}{5} r^4$ are represented in Fig.\ref{PRESS}, together with the metric function of CKG, inner ($r/R<1$) and outer. For $\lambda >0$ a horizon is present at finite $r$, while for $\lambda <0$
it grows as $r^4$. The pressure lines exchange. The value $\lambda=0$ is GR.\\

\begin{figure}[h]
\begin{center}
\includegraphics*[width=7cm,clip=]{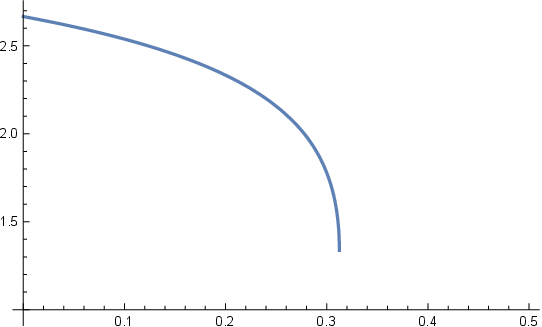}
\caption{\label{BUCH} The critical radius $\mu_m R^2$ as a function of $\lambda/\mu_m^2$. For $\lambda =0$ the value is $8/3$.
The lowest value is $\mu_m R^2=4/3$ for $\lambda/\mu_m^2 = 5/16$.}
\end{center}
\end{figure}

The analogue of the Buchdahl limit is now obtained, for $\Lambda =0$.\\ 
The right-hand-side of eq.\eqref{resultTOV} is zero at $r=0$ (i.e. the central pressure diverges) for:
\begin{align*} 
0 =
 \frac{1}{\mu_m} \left [\frac{\mu^2_m}{36}+\frac{\lambda}{5}\right ]- \frac{1}{4}  \left [\frac{\mu_m}{6}+\frac{\lambda}{5}R^2
 \right ]  + \frac{\mu_m}{24}  \sqrt{ 1 - \frac{2M(R)}{R}}
\end{align*}
With $X=\frac{1}{3}\mu_m R^2$, $Y=\frac{\lambda }{5}R^4$, it is
$$0= -\frac{1}{6}+\frac{4}{3} \frac{Y}{X^2} - \frac{Y}{X} +\frac{1}{2}\sqrt{1-X-Y} $$
The equation is solved for $Y$. A perfect square appears in the square root of the solution:
$$Y=\frac{  \frac{4}{9}-\frac{X}{3}-\frac{X^2}{4} \pm \left( \frac{X^2}{4} -\frac{5X}{3} +\frac{4}{3} \right ) }{2\left ( \frac{4}{3X}-1\right )^2}
$$
The plus is chosen because when $\lambda =0$ (i.e. $Y=0$) the equation gives Buchdahl's result. Then:
$(\frac{4}{3}-X)^2 Y = X^2(\frac{8}{9}-X)$, i.e.
$$\frac{\lambda}{5}(\mu_m R^2)^2 - \left( \frac{8}{5}\lambda -\frac{\mu_m^2}{3}\right) (\mu_m R^2) +\frac{16}{5}\lambda - \frac{8}{9}\mu_m^2 =0$$
The solution   
\begin{align}
\mu_m R^2 = 4 - \frac{5}{6}\frac{\mu_m^2}{\lambda} \left [1- \sqrt{1-\frac{16}{5}\frac{\lambda}{\mu_m^2}}\right ]
\end{align}
is the critical radius for $\Lambda =0$ in terms of $\mu_m$ and $\lambda $. It is plotted in Fig.\ref{BUCH}. The small $\lambda $ expansion recovers Buchdahl's relation as leading term, $\mu_m R^2=8/3$ (for the homogeneous star $M(R)=\frac{1}{6}\mu_m R^3$, then $M_{cr}=\frac{4}{9}R$). 

\section{Conclusions}
In a covariant formulation we obtain the extension of the Tolman-Oppenheimer-Volkoff equation in conformal Killing gravity. It differs by the presence of the dark pressure, and of the dark energy density that enters in the evaluation of the mass term $M(r)$.\\
We then study the junction conditions for a matter sphere in a CKG vacuum: continuity of the metric tensors 
and of the extrinsic curvatures. The latter is expressible as a pressure condition at the surface. 
The formulation of the theory
as an extension of the Einstein equation with an extra term $K_{jl}$, rather that the original third order Harada equations, 
is essential for implementing the standard procedure.\\
The junction conditions are applied to
the case of a sphere with uniform matter density, with an external solution given by Harada's vacuum \eqref{HARADAVAC}. With the hypothesis that the dark parameters are continuous across the boundary, 
we obtain the critical mass density-radius relation that generalizes Buchdahl's relation in GR.

\section*{Appendix 1: solutions of eq.\eqref{EQUATION1}}
The general solution of the inhomogeneous equation \eqref{EQUATION1} with $\kappa_3r^2+\kappa_4\ge 0$ is 
$y(r) = C y_0 (r)+ y_P(r)$ where $C$ is a constant, $y_0$ solves the homogeneous equation, $y_P(r)=y_0 (r) v(r)$ is a particular solution:
$$y_0(r) =  \frac{\sqrt{\kappa_3 r^2 +\kappa_4}}{r}, \qquad   v(r) =\int dr \frac{1+\kappa_1 r^2 + \kappa_2 r^4}{(\kappa_3 r^2 + \kappa_4)^{3/2}} $$
There are three cases to evaluate.\\
1) $\kappa_3>0,\, \kappa_4>0$.\\
In the integral set $r=\sqrt{\frac{\kappa_4}{\kappa_3}}\;{\rm sh}\theta $, 
$y_0 = \sqrt{\kappa_3} \frac{{\rm ch}\theta}{{\rm sh}\theta}$.
\begin{align*}
&v(r)= 
\frac{1}{ \sqrt{\kappa_3}} \int d\theta  \frac{\frac{1}{\kappa_4}+ \frac{\kappa_1}{\kappa_3} {\rm sh}^2 \theta +\frac{\kappa_2 \kappa_4}{\kappa^2_3}{\rm sh}^4 \theta }{{\rm ch}^2\theta}\\
%=& \frac{1}{C \sqrt{\kappa_3}} \int d\theta\, \left[\frac{\kappa_1}{\kappa_3} -2\frac{\kappa_2\kappa_4}{\kappa_3^2} + \left( \frac{1}
%{\kappa_4}-\frac{\kappa_1}{\kappa_3} +\frac{\kappa_2\kappa_4}{\kappa_3^2}\right )\frac{1}{{\rm ch}^2\theta} +
%\frac{\kappa_2\kappa_4}{\kappa_3^2} {\rm ch}^2\theta\right ]\\
&= \frac{1}{ \sqrt{\kappa_3}}  \big [\left(\frac{\kappa_1}{\kappa_3} -\frac{3}{2}\frac{\kappa_2\kappa_4}{\kappa_3^2} \right)\theta + \left( \frac{1}{\kappa_4}-\frac{\kappa_1}{\kappa_3} +\frac{\kappa_2\kappa_4}{\kappa_3^2}\right )\frac{{\rm sh}\theta}{{\rm ch}\theta}\\
& \quad+\frac{\kappa_2\kappa_4}{4\kappa_3^2} {\rm sh}(2\theta) \big]
\end{align*}
\begin{align*}
 y_P(r) =& \left[ \frac{\kappa_1}{\kappa_3} -\frac{3}{2}\frac{\kappa_2\kappa_4}{\kappa_3^2} \right] 
 \theta\frac{{\rm ch}\theta}{{\rm sh}\theta}+ \left[ \frac{1}{\kappa_4}-\frac{\kappa_1}{\kappa_3} +\frac{\kappa_2\kappa_4}{\kappa_3^2}\right ]\\
 &+\frac{\kappa_2\kappa_4}{2\kappa_3^2} ({\rm sh}^2\theta+1) \\
 =& \left[\frac{\kappa_1}{\kappa_3} -\frac{3}{2}\frac{\kappa_2\kappa_4}{\kappa_3^2} \right]
 \left [  \frac{\sqrt{\kappa_3 r^2+\kappa_4}}{r\sqrt{\kappa_3}} {\rm Arsh}\sqrt{\frac{\kappa_3}{\kappa_4}}r -1\right ]\\
& +  \frac{1}{\kappa_4} +\frac{\kappa_2}{2\kappa_3}r^2
 \end{align*}
2) $\kappa_3>0$ and $\kappa_4<0$. \\
In the integral set $r= \sqrt{\frac{|\kappa_4|}{\kappa_3}}{\rm ch}\theta$,  
$y_0 = \sqrt{\kappa_3}\frac{ {\rm sh}\theta }{{\rm ch}\theta}$,
\begin{align*}
&v(r)=
 \frac{1}{\sqrt{\kappa_3}} \int d\theta  \frac{\frac{1}{|\kappa_4|}+ \frac{\kappa_1}{\kappa_3} {\rm ch}^2 \theta +\frac{\kappa_2 |\kappa_4|}{\kappa^2_3}{\rm ch}^4 \theta }{{\rm sh}^2\theta}\\
%=& \frac{1}{C \sqrt{\kappa_3}} \int d\theta\, \left[\frac{\kappa_1}{\kappa_3}+2\frac{\kappa_2 |\kappa_4|}{\kappa_3^2} + 
%\left( \frac{1}{|\kappa_4|}+\frac{\kappa_1}{\kappa_3} +\frac{\kappa_2|\kappa_4|}{\kappa_3^2}\right )\frac{1}{{\rm sh}^2\theta} 
%+\frac{\kappa_2|\kappa_4|}{\kappa_3^2} {\rm sh}^2\theta\right ]\\
%=& \frac{1}{ \sqrt{\kappa_3}}  \left[\left(\frac{\kappa_1}{\kappa_3}+\frac{3}{2}\frac{\kappa_2|\kappa_4|}{\kappa_3^2} \right)\theta - \left( \frac{1}{|\kappa_4|}+\frac{\kappa_1}{\kappa_3} +\frac{\kappa_2|\kappa_4|}{\kappa_3^2}\right )\frac{{\rm ch}\theta}{{\rm sh}\theta} +\frac{\kappa_2|\kappa_4|}{4\kappa_3^2} {\rm sh}(2\theta)\right ]\\
&= \frac{1}{ \sqrt{\kappa_3}}  \big[\left(\frac{\kappa_1}{\kappa_3}-\frac{3}{2}\frac{\kappa_2\kappa_4}{\kappa_3^2} \right)\theta + \left( \frac{1}{\kappa_4}-\frac{\kappa_1}{\kappa_3} +\frac{\kappa_2\kappa_4}{\kappa_3^2}\right )\frac{{\rm ch}\theta}{{\rm sh}\theta} \\
&\quad -\frac{\kappa_2\kappa_4}{4\kappa_3^2} {\rm sh}(2\theta)\big ]
\end{align*}
\begin{align*}
 &y_P(r) = \left(\frac{\kappa_1}{\kappa_3} -\frac{3}{2}\frac{\kappa_2\kappa_4}{\kappa_3^2} \right)\theta\frac{{\rm sh}\theta}{{\rm ch}\theta}+ \left( \frac{1}{\kappa_4}-\frac{\kappa_1}{\kappa_3} +\frac{\kappa_2\kappa_4}{\kappa_3^2}\right )
 \\
 &\qquad \quad -\frac{\kappa_2\kappa_4}{2\kappa_3^2} ({\rm ch}^2\theta-1) \\
 &= \left(\frac{\kappa_1}{\kappa_3} -\frac{3}{2}\frac{\kappa_2\kappa_4}{\kappa_3^2} \right)
 \left [  \frac{\sqrt{\kappa_3 r^2+\kappa_4}}{r\sqrt{\kappa_3}} {\rm Arch}\sqrt{\frac{\kappa_3}{|\kappa_4|}}r -1\right ]\\
 &\quad +  \frac{1}{\kappa_4} +\frac{\kappa_2}{2\kappa_3}r^2
 \end{align*}
3) $\kappa_3<0$ and $\kappa_4>0$ ($r^2\le \kappa_4/|\kappa_3|$).\\ 
In the integral set $r=\sqrt{\frac{\kappa_4}{|\kappa_3|}} \sin\theta $, $y_0 =  \sqrt{|\kappa_3|}\frac{ \cos\theta}{\sin\theta}$.
\begin{align*}
&v(r)= \frac{1}{ \sqrt{|\kappa_3|}} \int d\theta  \frac{\frac{1}{\kappa_4}+ \frac{\kappa_1}{|\kappa_3|} {\rm sin}^2 \theta +\frac{\kappa_2 \kappa_4}{\kappa^2_3}{\rm sin}^4 \theta }{{\rm cos}^2\theta}\\
%=& \frac{1}{C \sqrt{\kappa_3}} \int d\theta\, \left[\frac{\kappa_1}{\kappa_3}+2\frac{\kappa_2 |\kappa_4|}{\kappa_3^2} + 
%\left( \frac{1}{|\kappa_4|}+\frac{\kappa_1}{\kappa_3} +\frac{\kappa_2|\kappa_4|}{\kappa_3^2}\right )\frac{1}{{\rm sh}^2\theta} 
%+\frac{\kappa_2|\kappa_4|}{\kappa_3^2} {\rm sh}^2\theta\right ]\\
%=& \frac{1}{C \sqrt{|\kappa_3|}}  \left[\left(-\frac{\kappa_1}{|\kappa_3|}-\frac{3}{2}\frac{\kappa_2\kappa_4}{\kappa_3^2} \right)\theta +\left( \frac{1}{\kappa_4}+\frac{\kappa_1}{|\kappa_3|} +\frac{\kappa_2\kappa_4}{\kappa_3^2}\right )\frac{\sin\theta}{\cos\theta} +\frac{\kappa_2\kappa_4}{4\kappa_3^2} {\sin}(2\theta)\right ]\\
&= \frac{1}{\sqrt{\kappa_3}}  \big [\left ( \frac{\kappa_1}{\kappa_3}
- \frac{3}{2}\frac{\kappa_2\kappa_4}{\kappa_3^2} \right) \theta 
+ \left( \frac{1}{\kappa_4} -\frac{\kappa_1}{\kappa_3}+\frac{\kappa_2\kappa_4}{\kappa_3^2}\right )\frac{\sin\theta}{\cos\theta} \\
&\quad +\frac{\kappa_2\kappa_4}{4\kappa_3^2} \sin (2\theta)\big ]
\end{align*}
\begin{align*}
 y_P(r) = \left(\frac{\kappa_1}{\kappa_3} -\frac{3}{2}\frac{\kappa_2\kappa_4}{\kappa_3^2} \right)\theta\frac{\cos\theta}{\sin\theta}+ \left( \frac{1}{\kappa_4}-\frac{\kappa_1}{\kappa_3} +\frac{\kappa_2\kappa_4}{\kappa_3^2}\right )\\
 +\frac{\kappa_2\kappa_4}{2\kappa_3^2} (1-\sin^2\theta ) \\
 = \left(\frac{\kappa_1}{\kappa_3} -\frac{3}{2}\frac{\kappa_2\kappa_4}{\kappa_3^2} \right)
 \left [  \frac{\sqrt{\kappa_3 r^2+\kappa_4}}{r\sqrt{|\kappa_3|}} {\rm ArcSin}\sqrt{\frac{|\kappa_3|}{\kappa_4}}r -1\right ]\\
 +  \frac{1}{\kappa_4} +\frac{\kappa_2}{2\kappa_3}r^2
 \end{align*}
 
 \vfill
\section*{Appendix 2}
With $F(r) = 1-ar^2 -br^4$, the differential equation for $y$ is: $y' F + \frac{1}{2}y F' + \frac{r}{2} y^2=0$.\\
Divide by $y^2 F^{3/2}$, the equation now is:
$$  \frac{d}{dr}\left(\frac{1}{y\sqrt F} \right ) = \frac{1}{2} \frac{r}{F^{3/2} }$$
An integration gives
\begin{align*} 
\frac{1}{y(r)} = & \sqrt {F(r)} \left [ C - \frac{1}{2} \int_r^R   \frac{r'dr'}{F(r')^{3/2}} \right ] \\
%=& \sqrt{F(r)}  \left [ C - \frac{1}{4} \int_{r^2}^{R^2}   \frac{ dx }{ \left ( 1- ax - bx^2 \right )^{3/2} }\right  ] \\
=& \sqrt{F(r)}  \left [ C - \frac{1}{2(a^2+4b)}\left( \frac{a+2bR^2}{\sqrt {F(R)}} - \frac{a+2br^2}{\sqrt {F(r)}} \right )\right ]
\end{align*}
The boundary condition $p_m(R)=0$ fixes the constant $C$:
\begin{align*} 
\frac{1}{y(r)} =
 \sqrt{\frac{F(r)}{F(R)} } \left [ \frac{1}{\mu_m} - \frac{a+2bR^2}{2(a^2+4b)} \right ] +   \frac{a+2br^2}{2(a^2+4b)} 
\end{align*}
which is eq.\eqref{resultTOV}.

\end{document}